\documentclass[letter,scriptaddress,onecolumn, nofootinbib,prl,showkeys,12pt]{revtex4}
\usepackage{amssymb}
\usepackage{amsmath}
\usepackage{makeidx}
\usepackage{amsfonts}
\usepackage{graphicx,epstopdf}
\usepackage[ansinew]{inputenc}
\usepackage[usenames,dvipsnames]{pstricks}
\usepackage{subfigure}
\usepackage{pdfpages}
\usepackage{epsfig}
\usepackage{pst-grad}
\usepackage{pst-plot}
\usepackage[colorlinks,hyperindex]{hyperref}

\hypersetup{
colorlinks,
citecolor=blue,
linkcolor=black,
urlcolor=black,
}

\newcommand{\be}{\begin{equation}}
\newcommand{\ee}{\end{equation}}

\newcommand{\beq}{\begin{eqnarray}}
\newcommand{\eeq}{\end{eqnarray}}

\begin{document}

\title{Configurational Entropy for Travelling Solitons in Lorentz and CPT
Breaking Systems}
\author{R. A. C. Correa}
\email{rafael.couceiro@ufabc.edu.br}
\affiliation{  CCNH, Universidade Federal do ABC, 09210-580, Santo André, SP,
Brazil}
\author{Rold\~ao da Rocha}
\email{roldao.rocha@ufabc.edu.br}
\affiliation{CMCC, Universidade Federal do ABC, 09210-580, Santo André, SP,
Brazil}\affiliation{International School for Advanced Studies (SISSA), Via Bonomea 265, 34136 Trieste, Italy}
\author{A. de Souza Dutra}
\email{dutra@feg.unesp.br}
\affiliation{UNESP-Campus de Guaratinguetá, 12516-410, Guaratinguetá, SP,
Brazil}


\begin{abstract}
{\footnotesize{In this work we group three research topics apparently disconnected, namely 
solitons, Lorentz symmetry breaking and entropy. Following a recent work
[Phys. Lett. B \textbf{713} (2012) 304], we show that it is possible to
construct in the context of travelling wave solutions a configurational
entropy measure in functional space, from the field configurations. Thus, we
investigate the existence and properties of travelling solitons in Lorentz
and \textit{CPT} breaking scenarios for a class of models with two
interacting scalar fields. Here, we obtain a complete set of exact solutions
for the model studied which display both double and single-kink
configurations. In fact, such models are very important in applications that
include Bloch branes, Skyrmions, Yang-Mills, Q-balls, oscillons and various
superstring-motivated theories. We find that the so-called Configurational
Entropy (CE) for travelling solitons, which we name as travelling
Configurational Entropy (TCE), shows that the best value of parameter
responsible to break the Lorentz symmetry is one where the energy density is
distributed equally around the origin. In this way, the
information-theoretical measure of travelling solitons in Lorentz symmetry
violation scenarios opens a new window to probe situations where the
parameters responsible for breaking the symmetries are random. In this case,
the TCE selects the best value.}}
\end{abstract}

\maketitle

\section{1. Introduction}

The most fundamental symmetry of the standard model of particle physics is
the Lorentz invariance, which has been very well verified in several
experiments. However, the first possibility of the Lorentz symmetry breaking
was announced by Kostelecky and Samuel \cite{kostelecky1}. They argued that superstring theories indicate
that Lorentz symmetry should be violated at higher energies. After that
seminal work, a great number of works regarding the  Lorentz symmetry
violation (LSV) have appeared in the literature. Nowadays, the breaking of
the Lorentz symmetry is a prominent  mechanism for the description of several
problems and conflicts in many areas of physics, from astrophysical \cite%
{kostelecky3,kosteleccky4,corrol,bertoline1,bertoline2}  to subatomic scales \cite{kostelecky5,musad,carlos,strumia,kostelecky6,kostelecky7,kostelecky8}. For example, it was shown in an inflationary scenario with
LSV \cite{kanno} that, using a scalar-vector-tensor theory with Lorentz
violation, the exact Lorentz violation inflationary solutions are found
without the presence of the inflaton potential.

An important investigation line about topological
defects in the presence of LSV have been recently addressed in the literature \cite%
{barreto,dutra1,menezes,rafael1}. In this context, it was shown by Barreto and
collaborators \cite{barreto} that the violation of Lorentz and \textit{CPT}
symmetries is responsible by the appearance of an asymmetry between defects
and anti-defects. Thus, in that context the authors showed that an analogous
investigation can be used to build string theory scenarios. Motivated by
this result,  a class of travelling
solitons in Lorentz and \textit{CPT} breaking systems was presented in Ref. \cite{rafael1}, where the solutions
present a critical behavior controlled by the choice of an arbitrary
integration constant. In this case,  the field
configurations habe been shown to allow the emergence of so-called superluminal solitons \cite%
{aharonov}. Another increasing interest in LSV arises in investigations on
neutrinos \cite{neutrinos}, gravity \cite{gravity}, electrodynamics \cite{eletrodynamics}, acoustic black hole \cite{acousticblackhole, acoustic},
monopoles and vortices \cite{seifert,seifert1,barraz,manoel1,casana1,julio,manoel2}.

On the other hand, in 1948, in an apparently disconnected topic, Shannon 
\cite{shannon} described what is  called  ``A mathematical theory of
communication", which nowadays is known as ``Information theory". In that
work, Shannon introduced a mathematical theory capable of solving  the most
fundamental problem of communication, namely, the  information transmission
either exactly or approximately.

The main purpose of the information theory presented in \cite{shannon} was
to introduce the concepts of entropy and mutual information, by  using the
viewpoint of communication theory. In this context, the entropy was defined
as a measure of ``uncertainty" or ``randomness" of a random phenomenon.
Thus, if a little deal of information about a random variable is received,
the uncertainty decreases accordingly. As a consequence, one can measure this reduction
in the uncertainty, which can be related to the quantity of transmitted
information. This quantity is the so-called mutual information. After that
work, a vast number of communication systems have been widely analysed from
 the information theory viewpoint, where the various types of
information transmission can be studied under a unified model.

Moreover, in a cosmological scenario, Bardeen, Carter and Hawking \cite%
{hawking} have established the relationship between the laws of
thermodynamics and black holes. Some years later, Wald \cite{wald}, motivated
by the connection between black-hole and thermodynamics, has precisely 
defined the entropy for a self-gravitating system which contains a black
hole. The ideas applied in \cite{wald} follow those ones provided  by 
information theory.

Finally, in a very recent work \cite{PLBGleiser}, the concept of entropy
has been, once more, reintroduced in the literature. Notwithstanding, now with an approach
capable of taking into account the dynamical and the informational contents
of models with localized energy configurations. In that letter, using an
analogy to the Shannon's information entropy, the Configurational Entropy (CE) was constructed. It can be applied to several nonlinear
scalar field models featuring solutions with spatially-localized energy. As
pointed out in \cite{PLBGleiser}, the CE can resolve situations where the
energies of the configurations are degenerate. In this case, the
configurational entropy can be used to select the best configuration. The
approach presented in \cite{PLBGleiser} have been used to study the
non-equilibrium dynamics of spontaneous symmetry breaking \cite{PRDgleiser-stamatopoulos}, to obtain the stability bound for compact
objects \cite{PLBgleiser-sowinski}, to investigate the emergence of
localized objects during inflationary preheating \cite{PRDgleiser-graham},
and moreover to distinguish configurations with energy-degenerate spatial profiles as well 
\cite{PLBrafael-alvaro-gleiser}.

Hence, in this work we shall construct a CE in functional space, which we
name Travelling Configurational Entropy (TCE), to measure the information
of travelling solitons. It is worth to remark that the TCE can be used to
study any physical model with energy density localized described by a
travelling variable. In this case, the entropic measure opens a new
theoretical window to probe the ordered arrangement of structures such as
topological defects \cite{vachaspati}, ferromagnetic materials \cite{gunton}%
, solids far from equilibrium \cite{langer}, and cosmic string \cite%
{vilenkin} likewise. As an application, we shall investigate classical field theories
in the context of Lorentz symmetry breaking and \textit{CPT} violation, which
admit energy density localized solutions. The model \cite{rajaramam,boia,bazeia,bnrt,bazeiaaa,dutra,rus1,rus2,rus3} that we will analyze has two interacting scalar fields and admits a
variety of kink-like solutions. Such model  has been shown in the literature to give rise to Bloch branes \cite{blochbrane,blochbrane0,blochbrane1},
electrical conductivity phenomena in superconductors \cite{brito}, bags,
junctions, and in addition networks of BPS and non-BPS defects \cite{brito1, salamanca}.

This paper is organized as follows: in Section 2 we present the model which
is going to be analyzed and we find the classical field configurations associated to it. In
Section 3 the TCE measure is defined and we compute the information-entropic
for travelling solitons in Lorentz and \textit{CPT} breaking systems. In
Section 4 we present our conclusions and final remarks.

\section{2. The Model}

In this section we investigate classical field theories in the context of
Lorentz symmetry breaking and \textit{CPT} violation. In this case, the
framework to study Lorentz and \textit{CPT} violation is the so-called
Standard-Model Extension (SME). Thus, in this context, we consider a
two-field model in $(1+1)$ dimensions, where the Lorentz breaking Lagrangian
density generalizes some works in the literature. In our theory, the
Lagrangian density contains both vector functions and tensor terms. At this
point, it is important to remark that the vector functions, which have a
dependence on the dynamical scalar fields, are responsible for the Lorentz
symmetry breaking. On the other hand, the tensor term breaks the Lorentz
and, eventually, the \textit{CPT} symmetry. Our motivation to write down such
Lagrangian density comes from the work presented by Potting \cite{potting},
where scalar field-theoretic models were presented. Hence, our generalized
Lorentz breaking Lagrangian density is described by
\begin{equation}
\mathcal{L}=\frac{1}{2}\,\eta _{1}^{\mu \nu }\partial _{\mu }{\phi }%
\,\partial _{\nu }\phi +\frac{1}{2}\,\eta _{2}^{\mu \nu }\partial _{\mu }{%
\chi }\,\partial _{\nu }\chi +H^{\mu }\partial _{\mu }\phi +J^{\nu }\partial
_{\nu }\chi +\partial _{\mu }\phi \,K^{\mu \nu }\,\partial _{\nu }\chi
-V(\phi ,\chi )\,,  \label{1}
\end{equation}%
\noindent where $\mu ,\nu =0,1$; $V(\phi ,\chi )$ denotes the self-interaction
potential; $H^{\mu }$ and $J^{\nu }$ are arbitrary vector functions of the
fields $\phi $ and $\chi$. In this case, such vector functions are
responsible for the breaking of the Lorentz symmetry. On the other hand, $%
K^{\mu \nu }=K^{\mu \nu }(\phi ,\chi )$ is a tensor function which represents the
source of both LSV and \textit{CPT} breaking symmetry. Here $K^{\mu \nu }$ is
represented by a\ $2\times 2$ matrix written in the form 
\begin{equation}
K^{\mu \nu }=\left( 
\begin{array}{cc}
K^{00}(\phi ,\chi ) & K^{01}(\phi ,\chi ) \\ 
K^{10}(\phi ,\chi ) & K^{11}(\phi ,\chi )%
\end{array}%
\right) .
\end{equation}

At this point, it is important to remark that the above matrix has arbitrary
elements. However, if this matrix is real, symmetric, and traceless, the \textit{%
CPT }symmetry is kept \cite{matrix1, matrix2}. Recently, \ a great number of
works using a similar process for break the Lorentz symmetry, with a tensor
like $K^{\mu \nu }$, have been used in the literature, from microscope \cite%
{agostini} to cosmological scales \cite{acousticblackhole, acoustic}.

Moreover,  the two effective metrics in the Lagrangian density (\ref{1})  
can be thought of as being perturbations of the Minkowski metric $\eta ^{\mu
\nu }$: 
\begin{eqnarray}
\eta _{1}^{\mu \nu }(\phi ,\chi ) &=&\eta ^{\mu \nu }+F^{\mu \nu }(\phi
,\chi ), \\
\eta _{2}^{\mu \nu }(\phi ,\chi ) &=&\eta ^{\mu \nu }+G^{\mu \nu }(\phi
,\chi ),
\end{eqnarray}%
\noindent for arbitrary tensors components $F^{\mu \nu }$ and $G^{\mu \nu }$
with norm much less than the unity. The motivation to include such
perturbations in the metric comes from the fact that the coefficients for LSV
cannot be removed from the Lagrangian density, by using variables or fields
redefinitions. Thus, observable effects of the LSV can be detected in the
above theory.

Now, from the Lagrangian density (\ref{1}), the Euler-Lagrange equations of
motion for the two scalar fields $\phi $ and $\chi $ are respectively
provided by: 
\begin{eqnarray}
&&\left. \eta _{1}^{\mu \nu }\partial _{\mu }\,\partial _{\nu }\phi +K^{\mu
\nu }\,\partial _{\mu }\partial _{\nu }\chi +(\partial _{\mu }\eta _{1}^{\mu
\nu })\partial _{\nu }\phi +\partial _{\mu }H^{\mu }+(\partial _{\mu }K^{\mu
\nu })\partial _{\nu }\chi -\frac{1}{2}\eta _{1\phi }^{\mu \nu }\partial
_{\mu }\phi \partial _{\nu }\phi -\frac{1}{2}\eta _{2\phi }^{\mu \nu
}\partial _{\mu }\chi \partial _{\nu }\chi \right.  \notag \\
&&\qquad \left. -H_{\phi }^{\mu }\partial _{\mu }\phi -J_{\phi }^{\nu
}\partial _{\nu }\chi -\partial _{\mu }\phi \,K_{\phi }^{\mu \nu }\,\partial
_{\nu }\chi +V_{\phi }=0\,,\right.  \label{eta1}
\end{eqnarray}%
\noindent and 
\begin{eqnarray}
&&\left. \eta _{2}^{\mu \nu }\partial _{\mu }\,\partial _{\nu }\phi +K^{\mu
\nu }\,\partial _{\mu }\partial _{\nu }\chi +(\partial _{\nu }\eta _{2}^{\mu
\nu })\partial _{\mu }\phi +\partial _{\nu }J^{\nu }+(\partial _{\nu }K^{\mu
\nu })\partial _{\mu }\phi -\frac{1}{2}\eta _{1\chi }^{\mu \nu }\partial
_{\mu }\phi \partial _{\nu }\phi -\frac{1}{2}\eta _{2\chi }^{\mu \nu
}\partial _{\mu }\chi \partial _{\nu }\chi \right.  \notag \\
&&\qquad \left. -H_{\chi }^{\mu }\partial _{\mu }\phi -J_{\chi }^{\nu
}\partial _{\nu }\chi -\partial _{\mu }\phi \,K_{\chi }^{\mu \nu }\,\partial
_{\nu }\chi +V_{\chi }=0\,,\right.  \label{eta2}
\end{eqnarray}%
\noindent where $A_{\phi }\equiv \partial A/\partial _{\phi }$ [$A_{\chi
}\equiv \partial A/\partial _{\chi }$], for any quantity $A$ dependent on $%
\phi \;[\chi ]$. It can be seen that the two above equations carry 
information about the LSV of the model, through the presence of the tensors $%
K^{\mu \nu }$, the $\eta _{i}^{\mu \nu }$ metrics, and the vector functions as well.

For the sake of simplicity, and with our loss of generality, let us suppose
that 
\begin{eqnarray}
F^{\mu \nu }(\phi ,\chi ) &=&f^{\mu \nu }=const., \\
G^{\mu \nu }(\phi ,\chi ) &=&g^{\mu \nu }=const., \\
K^{\mu \nu }(\phi ,\chi ) &=&k^{\mu \nu }=const.,
\end{eqnarray}
where, $f^{\mu \nu }$, $g^{\mu \nu }$, and $k^{\mu \nu }$ are given
by the matrices
\begin{equation}
f^{\mu \nu }=\left( 
\begin{array}{cc}
f^{00} & f^{01} \\ 
f^{10} & f^{11}%
\end{array}%
\right)\,,\quad g^{\mu \nu }=\left( 
\begin{array}{cc}
g^{00} & g^{01} \\ 
g^{10} & g^{11}%
\end{array}%
\right)\,,\quad k^{\mu \nu }=\left( 
\begin{array}{cc}
k^{00} & k^{01} \\ 
k^{10} & k^{11}%
\end{array}%
\right) .
\end{equation}

In $(1+1)$ dimensions, Eqs. (\ref{eta1}) and (\ref{eta2}) can be
respectively expressed as 
\begin{eqnarray}
&&\left. \eta _{1}^{00}\ddot{\phi}+K^{00}\ddot{\phi}+\eta _{1}^{11}\phi
^{\prime \prime }+(\eta _{1}^{10}+\eta _{1}^{01})\dot{\phi}^{\prime
10}+K^{01})\dot{\chi}^{\prime }-H_{\phi }^{0}\dot{\phi}-J_{\phi }^{0}\dot{%
\chi}-H_{\phi }^{1}\phi ^{\prime }\right.  \notag \\
&&\qquad \qquad \left. -J_{\phi }^{1}\chi ^{\prime }+\dot{H}%
^{0}+H^{1^{\prime }}+V_{\phi }=0\,,\right.  \label{6} \\
&&  \notag \\
&&\left. \eta _{2}^{11}\chi ^{\prime \prime 00}\ddot{\phi}+\eta _{2}^{00}%
\ddot{\chi}+(\eta _{2}^{01}+\eta _{2}^{10})\dot{\chi}^{\prime 10}+K^{01})%
\dot{\phi}^{\prime }-H_{\chi }^{0}\dot{\phi}-J_{\chi }^{0}\dot{\chi}-H_{\chi
}^{1}\phi ^{\prime }\right.  \notag \\
&&\qquad \qquad \left. -J_{\chi }^{1}\phi ^{\prime }+\dot{J}%
^{0}+J^{1^{\prime }}+V_{\chi }=0\,,\right.  \label{7}
\end{eqnarray}%
\noindent where the prime [dot] stands for the derivative with respect to
the space [time] dimension.

In general, as a consequence of the model studied in this work, we cannot
analytically solve the above differential equations. However, one can still
consider an interesting case for the field configurations, where travelling wave solutions are searched for.  Travelling waves configurations have
an important impact when boundary states for D-branes and 
supergravity fields in a D-brane are regarded as well \cite{hikida,bachas,turton}. Hence, in order
to solve analytically the equations (\ref{6}) and (\ref{7}) we apply the
redefinition 
\begin{equation}
u=Ax+Bt\,.  \label{tv}
\end{equation}

Thus, the fields $\phi $ and $\chi $ take the form 
\begin{eqnarray}
\phi (x,t) &\mapsto &\phi (u), \\
\chi (x,t) &\mapsto &\chi (u).
\end{eqnarray}

Therefore, Eqs. (\ref{6}) and (\ref{7}) are respectively led to the following
expressions: 
\begin{eqnarray}
&&\left. [B^{2}\eta _{1}^{00}+A^{2}\eta _{1}^{11}+AB(\eta _{1}^{10}+\eta
_{1}^{01})]\phi _{uu}+[B^{2}\,K^{00}+A^{2}\,K^{11}+AB(K^{10}+K^{01})]\chi
_{uu}\right.  \notag \\
&&\left. +[BH_{\phi }^{0}+AH_{\phi }^{1}-BH_{\phi }^{0}-AH_{\phi }^{1}]\phi
_{u}+[BH_{\chi }^{0}+AH_{\chi }^{1}-BJ_{\phi }^{0}-AJ_{\phi }^{1}]\chi
_{u}+V_{\phi }=0,\right.  \label{9}
\end{eqnarray}%
\noindent and%
\begin{eqnarray}
&&\left. [B^{2}\,K^{00}+A^{2}\,K^{11}+AB(K^{10}+K^{01})]\phi
_{uu}+[B^{2}\,\eta _{2}^{00}+A^{2}\,\eta _{2}^{11}+AB(\eta _{2}^{10}+\eta
_{2}^{01})]\chi _{uu}\right.  \notag \\
{}
&&\left. +[BJ_{\phi }^{0}+AJ_{\phi }^{1}-BH_{\chi }^{0}-AH_{\chi }^{1}]\phi
_{u}+[BJ_{\chi }^{0}+AJ_{\chi }^{1}-BJ_{\chi }^{0}-AJ_{\chi }^{1}]\chi
_{u}+V_{\chi }=0\,.\right.  \label{10}
\end{eqnarray}%
\noindent Such system of coupled equations can be simplified and further
computed by naming 
\begin{eqnarray}
\alpha := &&-B^{2}\eta _{1}^{00}-A^{2}\eta _{1}^{11}-AB(\eta _{1}^{10}+\eta
_{1}^{01}),  \label{alpha} \\
{}
\beta := &&B^{2}\,K^{00}+A^{2}\,K^{11}+AB(K^{10}+K^{01}),  \label{beta} \\
{}
\gamma := &&-B^{2}\,\eta _{2}^{00}-A^{2}\,\eta _{2}^{11}+AB(\eta
_{2}^{10}+\eta _{2}^{01}),  \label{gamma} \\
{}
L:= &&BH_{\phi }^{0}+AH_{\phi }^{1}-BH_{\phi }^{0}-AH_{\phi }^{1},  \label{L}
\\
{}
S:= &&BH_{\chi }^{0}+AH_{\chi }^{1}-BJ_{\phi }^{0}-AJ_{\phi }^{1},  \label{S}
\end{eqnarray}%
\noindent Hence by denoting $A_{u}\equiv \partial A/\partial u$ for any
quantity $A$, the system (\ref{9}) and (\ref{10}) can be expressed forthwith
as: 
\begin{eqnarray}
-\alpha \phi _{uu}+\beta \chi _{uu}+S\chi _{u}+V_{\phi } &=&0,  \label{19} \\
{}
\beta \phi _{uu}-\gamma \chi _{uu}-S\phi _{u}+V_{\chi } &=&0,  \label{20}
\end{eqnarray}%
\noindent which can be led to 
\begin{eqnarray}
-\alpha \phi _{u}\phi _{uu}+\beta \phi _{u}\chi _{uu}+S\phi _{u}\chi
_{u}+\phi _{u}V_{\phi } &=&0,  \label{21} \\
{}
\beta \chi _{u}\phi _{uu}-\gamma \chi _{u}\chi _{uu}-S\phi _{u}\chi
_{u}+\chi _{u}V_{\chi } &=&0,  \label{22}
\end{eqnarray}%
\noindent whose sum reads: 
\begin{equation}
-\alpha \phi _{u}\phi _{uu}-\gamma \chi _{u}\chi _{uu}+\beta (\phi _{u}\chi
_{uu}+\chi _{u}\phi _{uu})+\phi _{u}V_{\phi }+\chi _{u}V_{\chi }=0\,.
\end{equation}%
It implies that 
\begin{equation}
-\frac{\alpha }{2}\phi _{u}^{2}-\frac{\chi }{2}\chi _{u}^{2}+\beta \phi
_{u}\chi _{u}+V(\phi ,\chi )=E_{0}\,,
\end{equation}%
\noindent where $E_{0}$ is a constant of integration which can be lead to be
zero, in order to get solitonic solutions.

Now, the rescaling 
\begin{eqnarray}
\phi (u) &\mapsto &\sqrt{\alpha }\phi (u):=\tilde{\phi}(u), \\
{}
\chi (u) &\mapsto &\sqrt{\gamma }\chi (u):=\tilde{\chi}(u),
\end{eqnarray}%
\noindent makes Eq.(\ref{24}) to be written as 
\begin{equation}
-\frac{\tilde{\phi}_{u}^{2}}{2}-\frac{\tilde{\chi}}{2}+\frac{\beta }{\alpha
\gamma }\tilde{\phi}_{u}\tilde{\chi}_{u}+V(\tilde{\phi},\tilde{\chi})=0\,.
\label{24}
\end{equation}%
\noindent By rotating the variables $\tilde\phi$ and $\tilde\xi$ 
\begin{equation}
\binom{\tilde{\phi}}{\tilde{\chi}}=\frac{1}{\sqrt{2}}%
\begin{pmatrix}
1 & -1 \\ 
1 & 1%
\end{pmatrix}%
\binom{\theta }{\zeta },
\end{equation}%
\noindent and subsequently rescaling the new variables as 
\begin{equation}
\theta =\sqrt{\frac{2}{1+\tilde{\beta}}}\;\sigma \,,\qquad \qquad \quad
\zeta =\sqrt{\frac{2}{1-\tilde{\beta}}}\;\xi ,
\end{equation}%
we finally arrived at the following expression: 
\begin{equation}
-\frac{\sigma _{u}^{2}}{2}-\frac{\xi _{u}}{2}+V(\sigma ,\xi )=0\,.
\label{25}
\end{equation}%
\noindent It is worth to emphasize that the parameters $\alpha $ and $\gamma 
$ must be greater than zero, and it is necessary to impose for $\alpha >0$
that 
\begin{eqnarray}
B^{2}\eta _{1}^{00}+A^{2}\eta _{1}^{11}+AB(\eta _{1}^{10}+\eta _{1}^{01})
&<&0\,, \\
B^{2}\eta _{2}^{00}+A^{2}\eta _{2}^{11}+AB(\eta _{2}^{10}+\eta _{2}^{01})
&<&0\,.
\end{eqnarray}%
\noindent The parameters $\eta _{1}^{00}$ and $\eta _{2}^{00}$ can be thus
restricted by the other parameters that provide the Lorentz violation,
accordingly: 
\begin{eqnarray}
\eta _{1}^{00} &<&-\frac{A^{2}\eta _{1}^{11}+AB(\eta _{1}^{10}+\eta
_{1}^{01})}{B^{2}},  \label{33} \\
&&  \notag \\
\eta _{2}^{00} &<&-\frac{A^{2}\eta _{2}^{11}+AB(\eta _{2}^{10}+\eta
_{2}^{01})}{B^{2}}.  \label{34}
\end{eqnarray}%
The potential $V(\phi ,\chi )$ is supposed to be provided in terms of the
superpotential $W(\phi ,\chi )$, by 
\begin{equation}
V(\sigma ,\xi )=\frac{1}{2}{W_{\sigma }^{2}}+\frac{1}{2}{W_{\xi }^{2}},
\label{superr}
\end{equation}%
where $W_{\xi }={\partial W}/{\partial \xi }$ and $W_{\sigma }={\partial W}/{%
\partial \sigma }$. Notice that the critical points of the superpotential $%
W(\sigma ,\xi )$ provide the set of vacua $\{(\sigma ,\xi )\in \mathbb{R}%
^{2}:V(\sigma ,\xi )=0\}$ for the field theory model that is regarded. The
energy density has the form 
\begin{equation}
\epsilon (x)=\frac{1}{2}\left( \sigma ^{\prime \,2}+\xi ^{\prime
\,2}+W_{\sigma }^{2}+W_{\xi }^{2}\right) \,=\frac{1}{2}\left[ (\sigma
^{\prime }-W_{\sigma })^{2}+(\xi ^{\prime }-W_{\xi })^{2}\right] +dW\,.
\end{equation}%
The minimum energy solutions thus obey the expressions 
\begin{equation}
\sigma _{u}=\pm W_{\sigma },\hspace{0.5cm}\xi _{u}=\pm W_{\xi }\,,
\label{1sto}
\end{equation}%
leading us to the BPS energy \cite{bazeia}%
\begin{equation}
E_{\mathrm{BPS}}=\left\vert W(\sigma (\infty ),\xi (\infty ))-W(\sigma
(-\infty ),\xi (-\infty ))\right\vert ,  \label{ebps}
\end{equation}%
for smooth superpotentials. In terms of the superpotential, the equations of
motion for static fields read 
\begin{eqnarray}
\sigma ^{\prime \prime } &=&W_{\sigma }W_{\sigma \sigma }+W_{\xi }W_{\xi
\sigma }, \\
\xi ^{\prime \prime } &=&W_{\sigma }W_{\sigma \xi }+W_{\xi }W_{\xi \xi },
\end{eqnarray}%
which are solved by the first order equations \eqref{1sto}, for $W_{\sigma
\xi }=W_{\xi \sigma }$. Solutions to these first order equations are well
known to be BPS states, which solve the equations of motion. The sectors
where the potential has BPS states are named BPS sectors.

As an example, let us consider the model characterized by the superpotential 
\begin{equation}
W(\sigma ,\xi )=-\lambda \sigma +\frac{\lambda }{3}{\sigma ^{3}}-\mu \sigma
\xi ^{2}\,,  \label{lambda}
\end{equation}%
\noindent where $\lambda $ and $\mu $ are real positive dimensionless
coupling constants. The superpotential (\ref{lambda}) has been studied by
Shifman and Voloshin in the framework of $N=1$ supersymmetric Wess-Zumino
models with two chiral superfields \cite{shifman,shifmann}. In the purely
bosonic framework the presence of domain walls and its stability has been
analyzed in the references \cite{bazeia,bnrt,bazeiaaa}. Moreover, in \cite%
{M,M2} the complete structure of this type of solutions is given in two
critical values of the coupling between the two scalar fields, by exploiting
the integrability of the analogue mechanical system associated with this
model. This model, the so called BNRT model, has been further employed in
several systems in the context of field theory and condensed matter \cite%
{bnrt}.

For the superpotential (\ref{lambda}), the associated potential is provided
by 
\begin{equation}
V(\sigma ,\xi )=\frac{1}{2}\left[ \lambda ^{2}+\lambda ^{2}\sigma
^{2}(\sigma ^{2}-2)+\mu ^{2}\xi ^{2}\left( \xi ^{2}-\frac{2\lambda }{\mu }%
\right) +2\mu ^{2}\left( \frac{\lambda }{\mu }+2\right) \sigma ^{2}\xi ^{2}%
\right] .
\end{equation}%
\noindent For $\lambda /\mu >0$, the model presents four supersymmetric
minima $(\sigma ,\xi )$, given by: 
\begin{equation}
(\pm 1,0)\qquad \text{and}\qquad \left( 0,\sqrt{\frac{\lambda }{\mu }}%
\right) .
\end{equation}%
\noindent Now, from the set (\ref{1sto}) the following expression can be
derived: 
\begin{equation}
\frac{d\sigma }{d\xi }=\frac{W_{\sigma }}{W_{\xi }}=\frac{\lambda (\sigma
^{2}-1)+\mu \xi ^{2}}{2\mu \sigma \xi }.
\end{equation}%
\noindent Hence the following first order differential equation is
immediately derived: 
\begin{equation}
\frac{d\xi }{du}=%
\begin{cases}
\pm 2\mu \xi \sqrt{1+c_{0}\xi ^{\lambda /\mu }-\frac{\mu }{\lambda -2\mu }%
\xi ^{2}},\text{ }\qquad\lambda \neq 2\mu \,, \\ 
\\ 
\pm 2\mu \xi \sqrt{1+\xi ^{2}(\ln \xi +c_{1})},\text{ }\qquad \lambda =2\mu \,,%
\end{cases}
\label{46}
\end{equation}%
\noindent for $c_{0}$ and $c_{1}$ constants of integration. These equations
have analytical solutions. First, Eq.(\ref{46}) presents solutions, for $%
c_{0}<-2$ and $\lambda =\mu $: 
\begin{eqnarray}
\quad \xi ^{(1)}(u) &=&\frac{2}{\sqrt{c_{0}^{2}-4}\cosh (2\mu \,u)-c_{0}}\,, \\
&&  \notag \\
\quad \sigma ^{(1)}(u) &=&\frac{\sqrt{c_{0}^{2}-4}\sinh (2\mu \,u)}{\sqrt{%
c_{0}^{2}-4}\cosh (2\mu \,u)-c_{0}}\,.
\end{eqnarray}%
\noindent For $c_{0}<1/16$ and $\lambda =4\mu $: 
\begin{eqnarray}
\xi ^{(2)}(u) &=&-\frac{2}{\sqrt{\sqrt{1-16c_{0}}\cosh (4\mu \,u)+1}}, \\
&&  \notag \\
\quad \sigma ^{(2)}(u) &=&\frac{\sqrt{1-16c_{0}}\sinh (4\mu \,u)}{\sqrt{%
1-16c_{0}}\cosh (4\mu \,u)+1}.
\end{eqnarray}

These solutions for the corresponding original fields $\phi $ and $\chi$ thus read: 
\begin{eqnarray}
\phi ^{(j)}(u) &=&\frac{1}{\sqrt{\alpha }}\left( \frac{\sigma ^{(j)}(u)}{%
\sqrt{1+\tilde{\beta}}}-\frac{\xi ^{(j)}(u)}{\sqrt{1-\tilde{\beta}}}\right) ,
\\
&&  \notag \\
\chi ^{(j)}(u) &=&\frac{1}{\sqrt{\gamma }}\left( \frac{\sigma ^{(j)}(u)}{%
\sqrt{1+\tilde{\beta}}}+\frac{\xi ^{(j)}(u)}{\sqrt{1-\tilde{\beta}}}\right) ,
\end{eqnarray}%
\noindent for $j=1,2$.

The profile for the fields $\phi ^{(j)}(u)$ and $\chi ^{(j)}(u)$ are
depicted in Fig. 1, which shows the influence of the Lorentz violation on
the field configurations. Furthermore, in Fig. 2 we can see the orbits
connecting the vacua. In the next section we shall describe the so-called
configurational entropy (CE). In this case, similarly to the seminal result
by Gleiser and Stamatopoulos (GS) \cite{PLBGleiser}, we are going to
postulate the travelling configurational entropy, that can be employed in
order to analyze the entropic profile of any localized configuration of
fields. Besides, it can be further used in classical field theories
presenting solutions in the travelling variables.

\section{3. travelling Configurational Entropy (TCE)}

Recently GS showed that scalar field configurations, spatially localized and
with finite energy, presenting the same energy can be discriminated via the
so-called configurational entropy \cite{PLBGleiser}. Analogously to the
Shannon's information theory, the configurational entropy can be described
by the expression 
\begin{equation}
S_{c}[f]=-\int d^{d}\vec{k}\;\tilde{f}(\vec{k})\ln [\tilde{f}(\vec{k})]\,,
\label{sc}
\end{equation}\noindent where $d$ denotes the number of space dimensions, $\tilde{f}(%
\vec{k}):={f}(\vec{k})/{f}_{\mathrm{max}}(\vec{k})$, and ${f}(\vec{k})$ is
defined as the modal fraction 
\begin{equation}
{f}(\vec{k})=\frac{|F(\vec{k})|^{2}}{\int d^{d}\vec{k}\;|F(\vec{k})|^{2}}.
\label{53}
\end{equation}%
The quantity ${f}_{\mathrm{max}}(\vec{k})$ denotes the maximal modal
fraction, namely, the mode the contributes to the maximal contribution, and $%
F(\vec{k})$ was defined in \cite{PLBGleiser} as the Fourier transform of the
energy density. The higher the configurational entropy 
the higher the energy of the solutions, corresponding to the most
ordered solutions \cite{PLBrafael-alvaro-gleiser}. The configurational
entropy is moreover responsible to point out which solution is the most
ordered one among a family of infinite degenerated solutions. Hereon we
likewise propose that the function $F(\vec{k})$ also represents the Fourier
transform of the energy density as well. Notwithstanding, the variable of
integration is the travelling variable, namely 
\begin{equation}
F_{T}[\vec{k}]=\frac{1}{(2\pi )^{d/2}}\int d^{d}u\;e^{i\vec{k}\dot{\vec{u}}%
}\,T^{00}(\vec{u}).  \label{54}
\end{equation}%
\noindent Moreover, from the Plancherel theorem it follows that 
\begin{equation}
\int d^{d}u\;|F(\vec{k})|^{2}=\int d^{d}u|T^{00}(\vec{u})|^{2}.
\end{equation}

In this context, the modal fraction obeys the same relation (\ref{53}) and
the configurational entropy can be determined by Eq.(\ref{sc}). Thus we can
achieve the entropic measure of localized scalar configurations that present
their structure determined by the travelling variables. In this case, we
name the expression (\ref{sc}) as the Travelling Configurational Entropy
(TCE). Moreover, our framework can be straightforwardly led to the results
in \cite{PLBGleiser}, when the limit $B\rightarrow 0$ is taken in Eq.(\ref{tv}%
). The description heretofore presented can be hence applied to Lorentz
violation models in order to analyze the entropic profile of travelling
solitons. It is moreover worth to mention that the travelling-like solutions
can be recovered by adjusting the Lorentz violation parameters and leading
to the usual Lorentz symmetry.

In order to analyze the energy for the obtained configurations, the
energy-momentum tensor 
\begin{equation}
T^{\mu \nu }=\frac{\partial \mathcal{L}}{\partial (\partial _{\mu }\phi )}%
\partial ^{\nu }\phi +\frac{\partial \mathcal{L}}{\partial (\partial _{\mu
}\chi )}\partial ^{\nu }\chi -g^{\mu \nu }\mathcal{L},
\end{equation}%
\noindent is now regarded for the Lagrangian density: 
\begin{eqnarray}
&&\left. T^{00}(u)=\frac{1}{2}(\eta _{1}^{00}B^{2}+\eta _{1}^{11}A^{2})\phi
_{u}^{2}+\frac{1}{2}(\eta _{2}^{00}B^{2}+\eta _{2}^{11}A^{2})\chi
_{u}^{2}\right. \\
&&\left. \qquad\qquad+(K^{00}B^{2}+K^{11}A^{2})\chi _{u}\phi _{u}-A(H^{1}\phi
_{u}+J^{1}\chi _{u})+V(\phi ,\chi )\,.\right.
\end{eqnarray}

The profile of the energy density is presented in Fig. 3. 
Besides Eqs.(\ref{33}) and (\ref{34}), the constraint $-1<\tilde{\beta}<1$
holds, implying hence the following constraints for $K^{00}$: 
\begin{equation*}
-\frac{1}{B^{2}}\left[ \frac{\sqrt{\alpha \gamma }}{2}%
+A^{2}K^{11}+(K^{10}+K^{01})AB\right] <K^{00}<\frac{1}{B^{2}}\left[ \frac{%
\sqrt{\alpha \gamma }}{2}-A^{2}K^{11}-(K^{10}+K^{01})AB\right] .
\end{equation*}

In (1+1) dimensions, Eq.(\ref{54}) obviously reads 
\begin{equation}
F[{k}]=\frac{1}{\sqrt{2\pi }}\int_{-\infty }^{\infty }du\;e^{i{k}{u}%
}\,T^{00}({u}).
\end{equation}%
\noindent Now, by using Eq.(\ref{superr}) we get the following transform: 
\begin{equation}
F[{k}]=\frac{1}{\sqrt{2\pi }}\int_{-\infty }^{\infty }\;du\;e^{i{k}{u}}\left[
\frac{2b_{+}+1}{2}\sigma _{u}^{2}+\frac{2b_{-}+1}{2}\xi _{u}^{2}+b_{3}\sigma
_{u}\xi _{u}+b_{4}\sigma _{u}+b_{5}\xi _{u}\right] ,
\end{equation}%
\noindent where 
\begin{eqnarray}
b_{\pm }:= &&\frac{a_{1}^{2}}{2}\left( \frac{\gamma _{1}}{\alpha }+\frac{%
\gamma _{2}}{\gamma }\pm \frac{2\gamma _{3}}{\sqrt{\alpha \gamma }}\right) ,\qquad
b_{3}:= a_{1}a_{2}\left( \frac{\gamma _{1}}{\alpha }+\frac{\gamma _{2}}{%
\gamma }\right) ,\qquad  \\
&&  \notag \\
b_{4}:= &&-a_{1}\left( \frac{\gamma _{4}}{\sqrt{\alpha }}+\frac{\gamma _{5}}{%
\sqrt{\gamma }}\right) , \qquad \quad\quad 
b_{5}:= a_{2}\left( \frac{\gamma _{4}}{\sqrt{\alpha }}-\frac{\gamma _{5}}{%
\sqrt{\gamma }}\right) ,
\end{eqnarray}%
\noindent and 
\begin{eqnarray*}
&&\left. a_{1}=(1+\tilde{\beta})^{-1/2},\;a_{2}=(1-\tilde{\beta}%
)^{-1/2},\;\right. \\
&&\left. \gamma _{1}=\eta _{1}^{00}B^{2}+\eta _{1}^{11}A^{2},\;\gamma
_{2}=\eta _{2}^{00}B^{2}+\eta _{2}^{11}A^{2},\right. \\
&&\left. \gamma _{3}=K^{00}B^{2}+K^{11}A^{2},\;\gamma _{4}=AH^{1},\text{ }%
\gamma _{5}=AJ^{1}.\right.
\end{eqnarray*}

Therefore the function $F[k]$ can be written as 
\begin{equation}
F[k]=\sum_{\ell =1}^{4}\sum_{m=1}^{3}r_{(m)}(\ell )\,I^{(m)}(\ell ),
\end{equation}%
\noindent where 
\begin{eqnarray}
&&\left. I^{(1)}(\ell )=2^{\ell }\sum_{n=1}^{2}\tilde{I}_{n}(\ell ),\right.
\label{63} \\
&&  \notag \\
&&\left. I^{(2)}(\ell )=2^{\ell }\sum_{n=1}^{2}\sum_{j=1}^{2}(-1)^{n-1}%
\tilde{I}_{n}^{(\ell +1)}(\ell ),\right.  \label{64} \\
&&  \notag \\
&&\left. I^{(3)}(\ell )=2^{\ell +2}\sum_{n=1}^{2}\sum_{j=1}^{2}\bar{I}%
_{n}^{(j)}(\ell +3),\right.  \label{65}
\end{eqnarray}%
\noindent where $\bar{I}_{n}^{(j)}(\ell +3)=I_{n}^{(j)}(\ell +4)$. Besides, we have used the following notation: 
\begin{eqnarray}
&&\left. r_{(1)}(\ell )=q_{\ell },\qquad\text{ }r_{(2)}(\ell )=s_{\ell },\qquad\text{ }%
r_{(3)}(\ell )=p_{\ell }\,,\right. \\
&&\left. q_{1}=2\mu g_{4}c_{0},\qquad \text{ }q_{2}=-8\mu g_{4},\qquad \text{ }q_{3}=-8\mu
g_{4},\right. \\
&&\left. q_{4}=8\mu g_{1}-4\mu g_{2}(c_{0}^{2}-4),\text{ }s_{1}=-2g_{5}\sqrt{%
c_{0}^{2}-4}\,,\text{ }s_{2}=-4\mu g_{3}c_{0}\sqrt{c_{0}^{2}-4},\right. \\
&&\left. s_{3}=16\mu g_{3}\sqrt{c_{0}^{2}-4}\,,\text{ }\quad p_{1}=4\mu g_{2}\sqrt{%
c_{0}^{2}-4},\right.
\end{eqnarray}%
accordingly.

In Eqs.(\ref{63}-\ref{65}) it reads 
\begin{eqnarray}
&&\left. \tilde{I}_{n}(\ell )=\frac{1}{2\mu \lbrack
C_{0}(B_{0}^{2}-D_{0}^{2})]^{\ell }}\frac{\Gamma \left[ \ell +(-1)^{n-1}%
\frac{ik}{2\mu }\right] }{\Gamma \left[ \ell +1+(-1)^{n-1}\frac{ik}{2\mu }%
\right] }\times \right.  \notag \\
&&  \notag \\
&&\left. {}_2F_{1}\left[ \ell +(-1)^{n-1}\frac{ik}{2\mu },\ell ,\ell ,\ell
+1+(-1)^{n-1}\frac{ik}{2\mu };X_{1},Y_{1}\right] ,\right.  \label{68}
\end{eqnarray}%
\noindent and 
\begin{eqnarray}
&&\left. I_{n}(\ell )=\frac{1}{2\mu \lbrack
C_{0}(B_{0}^{2}-D_{0}^{2})]^{\ell }}\frac{\Gamma \left[ (-1)^{j-1}\bar{A}/%
\bar{B}+\ell +(-1)^{n+j}ik/\bar{B}\right] }{\Gamma \left[ (-1)^{j-1}\bar{A}/%
\bar{B}+\ell +1+(-1)^{n+j}ik/\bar{B}\right] }\times \right.  \notag \\
&&  \notag \\
&&\left. {}_2F_{1}\left[ (-1)^{j-1}\frac{\bar{A}}{\bar{B}}+\ell +(-1)^{n+j}\frac{%
ik}{\bar{B}},\ell ,\ell ,(-1)^{j-1}\frac{A}{B}+\ell +1+(-1)^{j+n}\frac{ik}{%
\bar{B}};X_{1},Y_{1}\right] ,\right.  \label{70}
\end{eqnarray}%
\noindent where $\bar{A}=2\mu $, $C_{0}:=\sqrt{c_{0}^{2}-4}$, $D_{0}:=\sqrt{%
c_{0}^{2}-C_{0}^{2}}/{C_{0}}$, $B_{0}:=c_{0}/C_{0}$, $X_{1}:=1/(B_{0}+D_{0})$%
, and $Y_{1}:=1/(B_{0}-D_{0})$.

Moreover, 
\begin{equation}
\bar{I}_{n}^{(j)}(\ell )={I}_{n}^{(j)}(\ell +1),
\end{equation}%
\noindent with $A=4\mu $ and $B=2\mu $.

The fraction mode can be thus written in a more compact form: 
\begin{equation}
f(k)=\frac{\sum_{\ell ,\ell ^{\prime }=1}^{4}\sum_{m,m^{\prime
}=1}^{3}\;r_{(m)}(\ell )\,r_{(m^{\prime })}(\ell ^{\prime })^{\ast
}I^{(m)}(\ell )\,I^{(m^{\prime })}(\ell ^{\prime })^{\ast }}{\sum_{\ell
,\ell ^{\prime }=1}^{4}\sum_{m,m^{\prime }=1}^{3}\int_{-\infty }^{\infty
}\,dk\;r_{(m)}(\ell )\,r_{(m^{\prime })}(\ell ^{\prime })^{\ast
}I^{(m)}(\ell )\,I^{(m^{\prime })}(\ell ^{\prime })^{\ast }}.
\end{equation}

In Fig. 4 we can realize the behavior of the modal fraction and realize how
its profile is influenced by the Lorentz violation parameters.

To compute the configurational entropy, we must integrate Eq. (\ref{sc})
numerically. The results are shown in Fig. 5, where the TCE is plotted as a
function of the parameter $k^{00}$. From that figure we can check that there
is a region of $k^{00}$ where the existence of solutions is forbidden by
entropy. In this case, the region of parameter travelling where the fields is
most prominent are that given by values $k^{00}>-0.06$ with the
corresponding TCE $S_{c}=0$. Moreover, for $k^{00}=-0.06$ the field
configurations undergo a kind of phase transition, where the two-kink
solution in the fields $\phi ^{(1)}(u)$ and $\chi ^{(1)}(u)$ converges into
a single kink. Another very important revelation that comes from entropy
measure has risen when $k^{00}\rightarrow -0.055$. In this limit we have a
symmetric distribution of energy density around the origin, showing that the
field configurations are equally distributed in both the sides of vertical
axes. Moreover, in that limit once more the configurations undergo a
new transition in their structures, where the associated  solutions converge in lumps
configurations.

For the sake of completeness, we have examined how the results vary with respect to the
fields $\phi ^{(2)}(u)$ and $\chi ^{(2)}(u)$, where we conclude that the
qualitative features remain the same.

The above results lead us to conclude that the TCE can be used in other to
extract a rich information about the structure of the configurations which
is clearly related to their travelling profiles. Here, we found that the best
ordering for the solutions are that given by $k^{00}\rightarrow -0.055$
where the configuration is symmetric around of origin.

\section{5. Conclusions}

In this work, following a recent work \cite{PLBGleiser}, we showed that it is
possible to construct a configurational entropy measure in functional space
from the field configurations where travelling wave solutions can be studied.
Thus, we applied the approach to investigate the existence and properties of
travelling solitons in Lorentz and \textit{CPT} breaking scenarios for a
class of models with two interacting scalar fields. Here, we obtained a
complete set of exact solutions for the model studied,  which display both
double and single-kink configurations. We have found that the so-called
Configurational Entropy for travelling solitons, which we name as
 the Travelling Configurational Entropy (TCE), shows that the best value for the 
parameter responsible to break the Lorentz symmetry is the one which has energy
density profile symmetric with respect to the origin. In this way, the
information-theoretical measure of travelling solitons in Lorentz symmetry
violation scenarios opens a new window to probe situations where the
parameters responsible for breaking the symmetries are random. In this case,
the TCE selects the best value. Moreover, the variable used in this work, 
$u=Ax+Bt$, when compared with the usual boosted variable, $u=\gamma (x+vt)$,
allows that the parameters $A$ and $B$ can be chosen in a range larger than the
corresponding ones in the boosted variable, allowing the appearance of
superluminal solitons \cite{aharonov}. Thus, the TCE provides a
complementary perspective to investigate the causality and superluminal
behavior in classical field theories such as k-essence theories and
MOND-like theories of gravity \cite{mond, kessence}. Other applications
where TCE can be used to relate the dynamical and informational content of
physical system is found in the so-called Galilean field theories. In this
context, we are presently interested in the possibility of constructing the
entropic profile of Galileons on cosmological backgrounds \cite{trodden1}.

\subsection*{Acknowledgements}

RACC thanks to UFABC and CAPES for financial support. RdR thanks to SISSA
for the hospitality and to CNPq grants No. 303027/2012-6 and No. 
473326/2013-2 for partial financial support. RdR is also supported in part
by the CAPES\emph{\ }Proc. 10942/13-0.\emph{\ }ASD was supported in part by
the CNPq. \ RACC also thanks to the Professor M. Gleiser for introducing him
to this matter.

\newpage

\begin{figure}[h]
\begin{center}
\includegraphics[width=12cm]{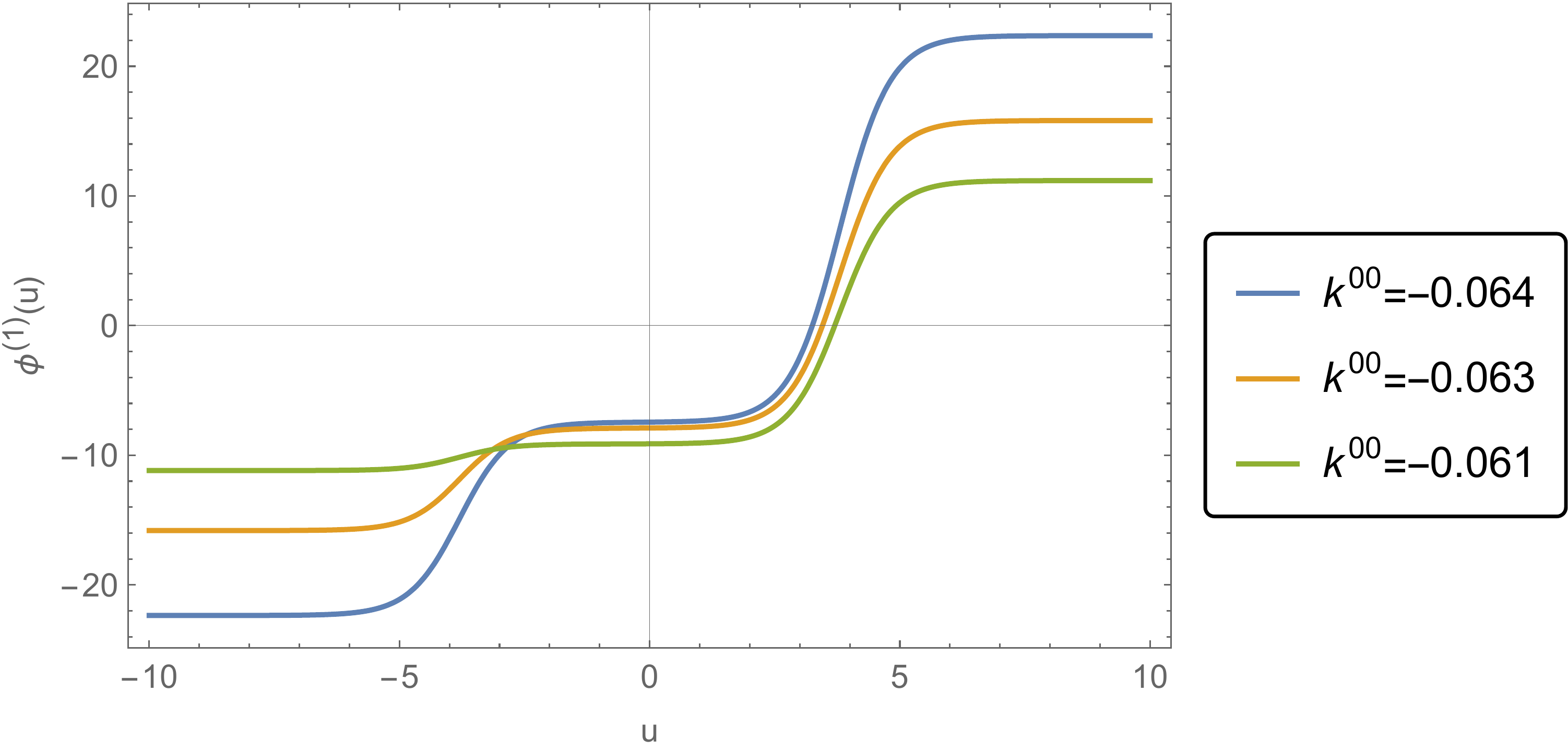} %
\includegraphics[width=12cm]{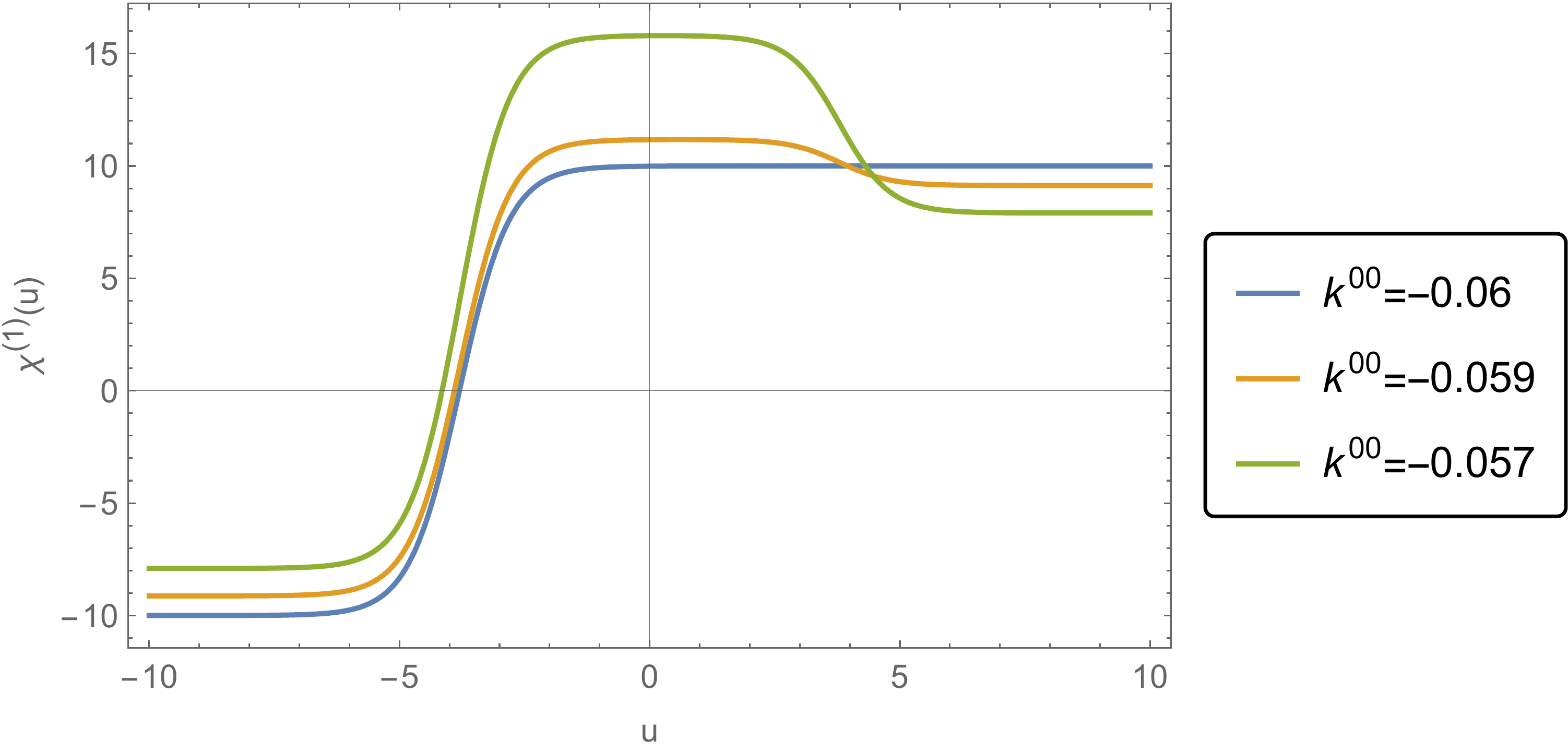}
\end{center}
\caption{Field configurations $\protect\phi ^{(1)}(u)$ and $\protect\chi %
^{(1)}(u)$ for $A=B=1,$ $\protect\mu =1,$ $\protect\eta _{1}^{00}=\protect%
\eta _{2}^{00}=-3.05,$ $\protect\eta _{1}^{11}=\protect\eta _{2}^{11}=1.01,$ 
$\protect\eta _{1}^{01}=\protect\eta _{2}^{01}=1.01,$ $\protect\eta %
_{1}^{10}=\protect\eta _{2}^{10}=1.02,$ $k^{01}=0.03,$ $k^{10}=0.02,$ $%
k^{11}=0.01,$ and $c_{0}=-2.000001.$}
\label{fig1:field configurations}
\end{figure}

\begin{figure}[h]
\begin{center}
\includegraphics[width=8.5cm]{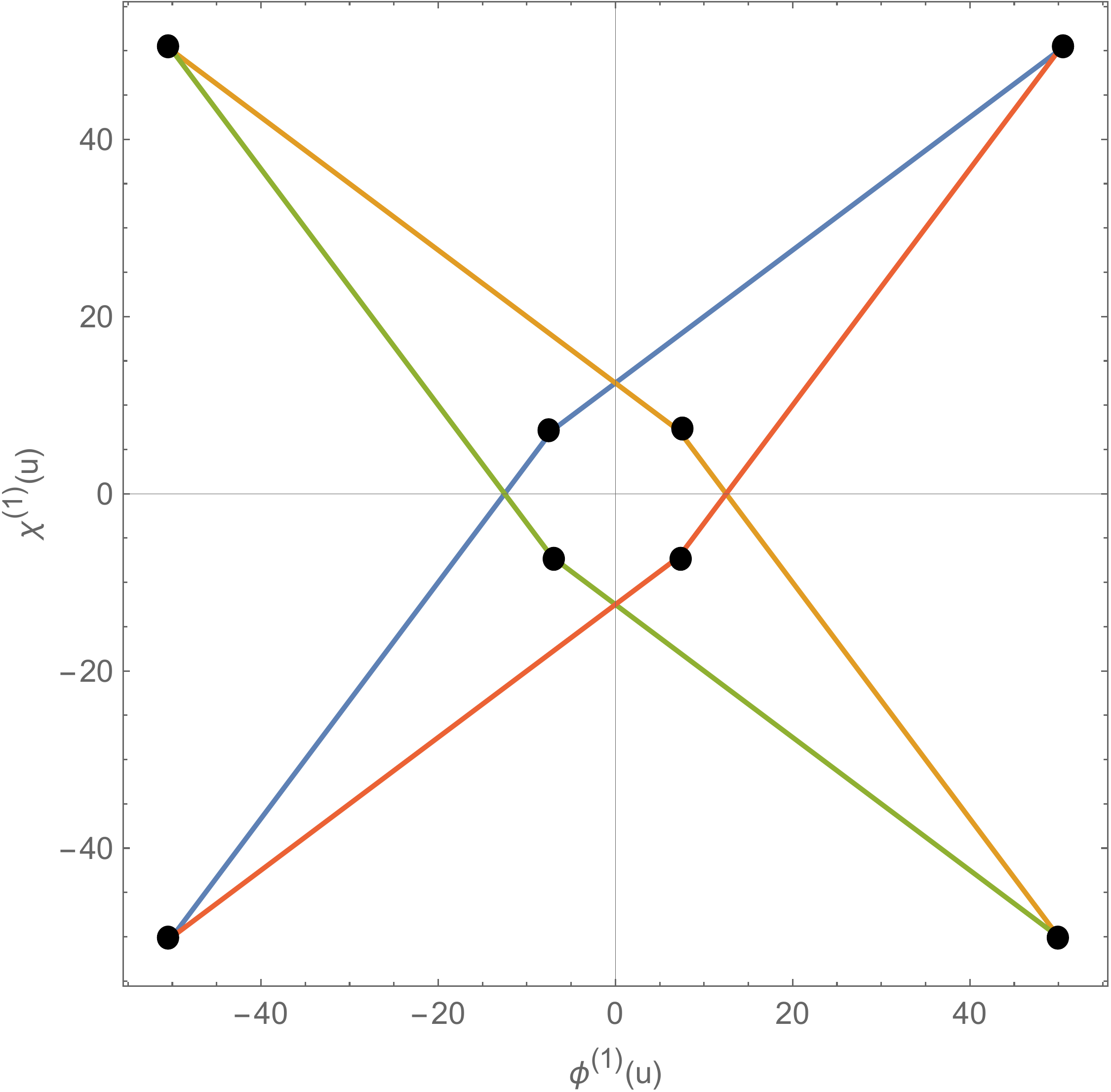} %
\includegraphics[width=8.5cm]{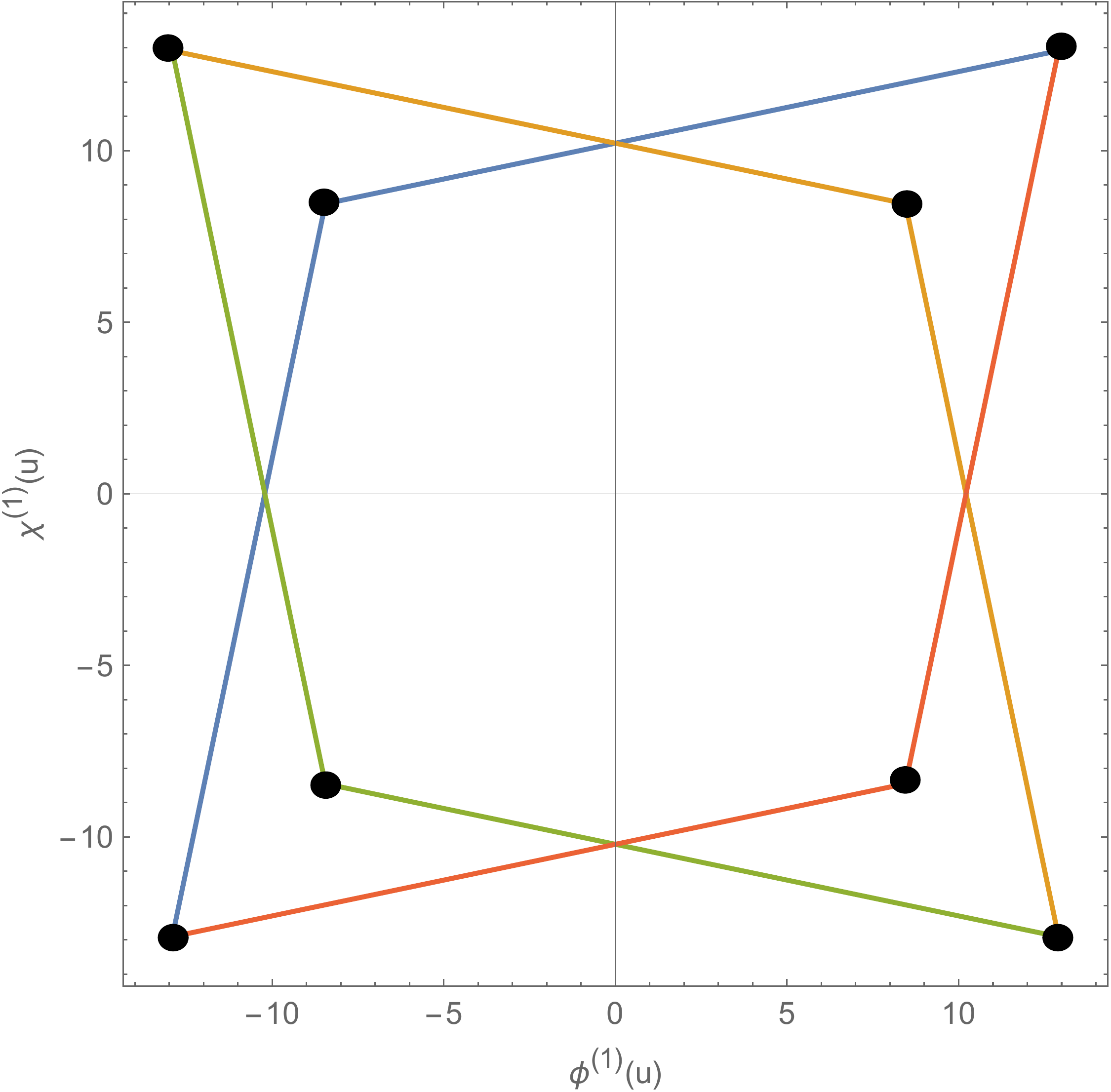}
\end{center}
\caption{Orbit for the solutions and vacuum states of the potential for $%
A=B=1,$ $\protect\mu =1,$ $\protect\eta _{1}^{00}=\protect\eta %
_{2}^{00}=-3.05,$ $\protect\eta _{1}^{11}=\protect\eta _{2}^{11}=1.01,$ $%
\protect\eta _{1}^{01}=\protect\eta _{2}^{01}=1.01,$ $\protect\eta _{1}^{10}=%
\protect\eta _{2}^{10}=1.02,$ $k^{01}=0.03,$ $k^{10}=0.02,$ $k^{11}=0.01,$
and $c_{0}=-2.000001$. The plot on the top of the figure show the case with $%
k^{00}=-0.0648$ and the bottom ones with $k^{00}=-0.0620$.}
\end{figure}

\begin{figure}[h]
\begin{center}
\includegraphics[width=12cm]{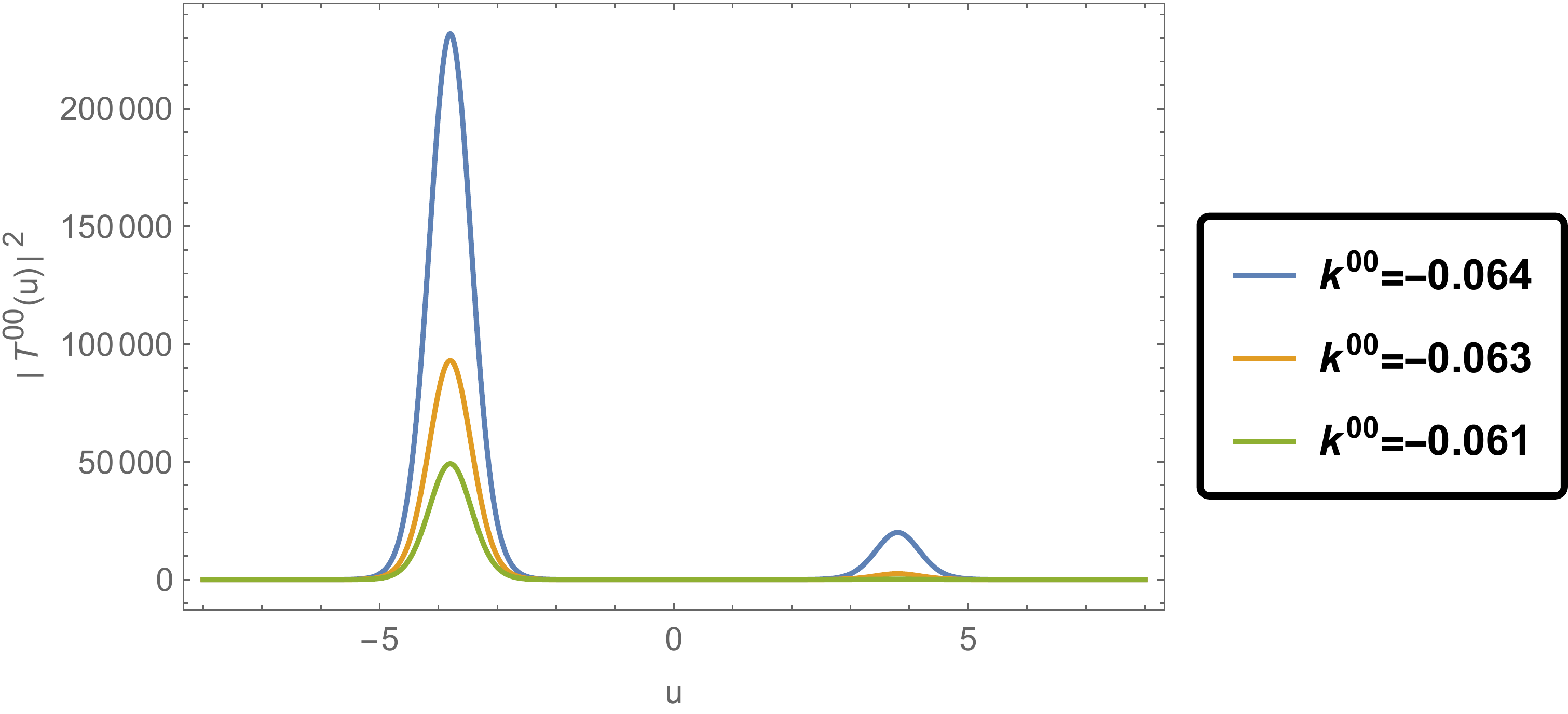}
\end{center}
\caption{Energy density for $A=B=1,$ $\protect\mu =1,$ $\protect\eta %
_{1}^{00}=\protect\eta _{2}^{00}=-3.05,$ $\protect\eta _{1}^{11}=\protect%
\eta _{2}^{11}=1.01,$ $\protect\eta _{1}^{01}=\protect\eta _{2}^{01}=1.01,$ $%
\protect\eta _{1}^{10}=\protect\eta _{2}^{10}=1.02,$ $k^{01}=0.03,$ $%
k^{10}=0.02,$ $k^{11}=0.01,$ and $c_{0}=-2.000001$. }
\end{figure}

\begin{figure}[h]
\begin{center}
\includegraphics[width=12cm]{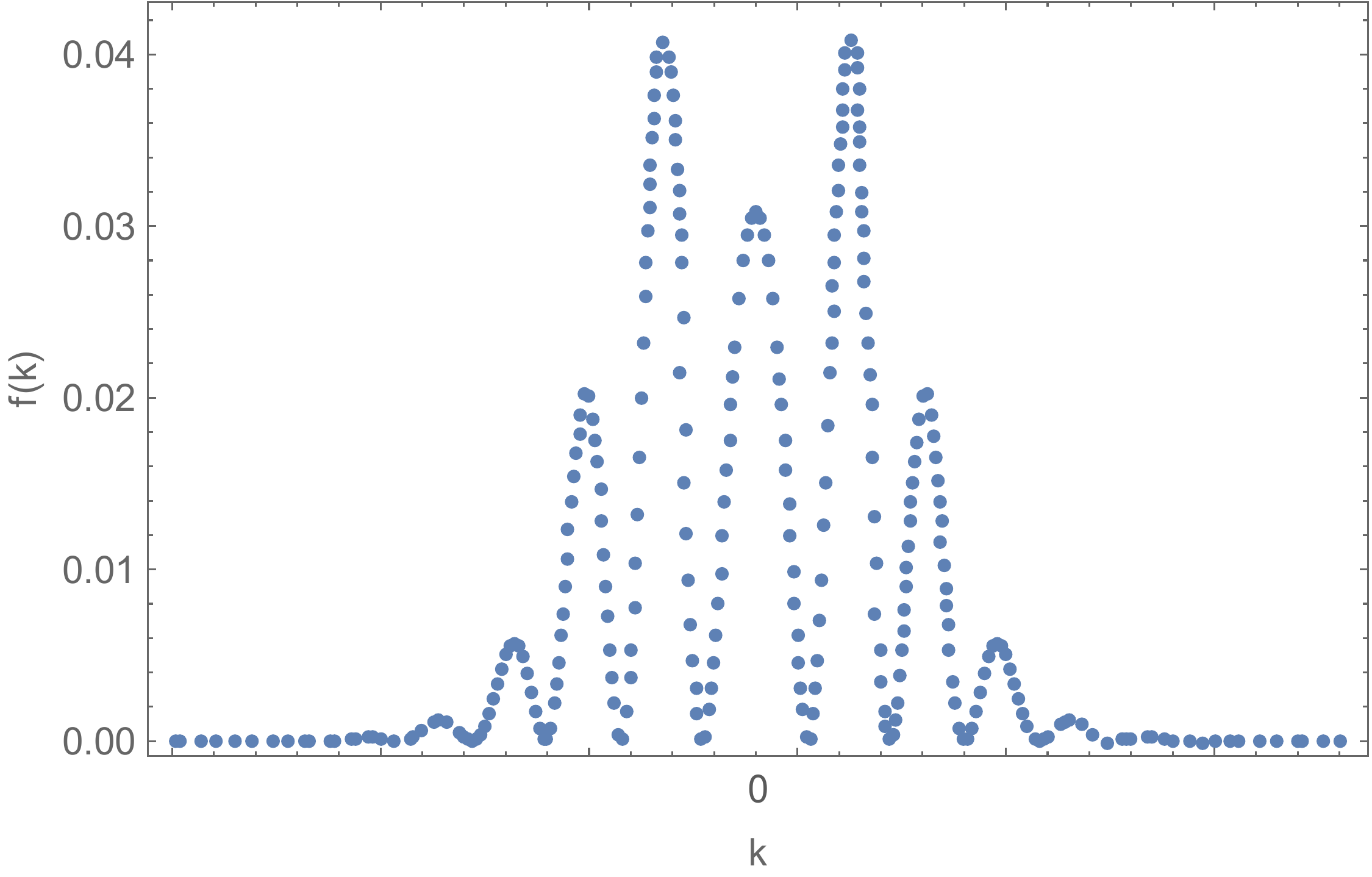} %
\includegraphics[width=12cm]{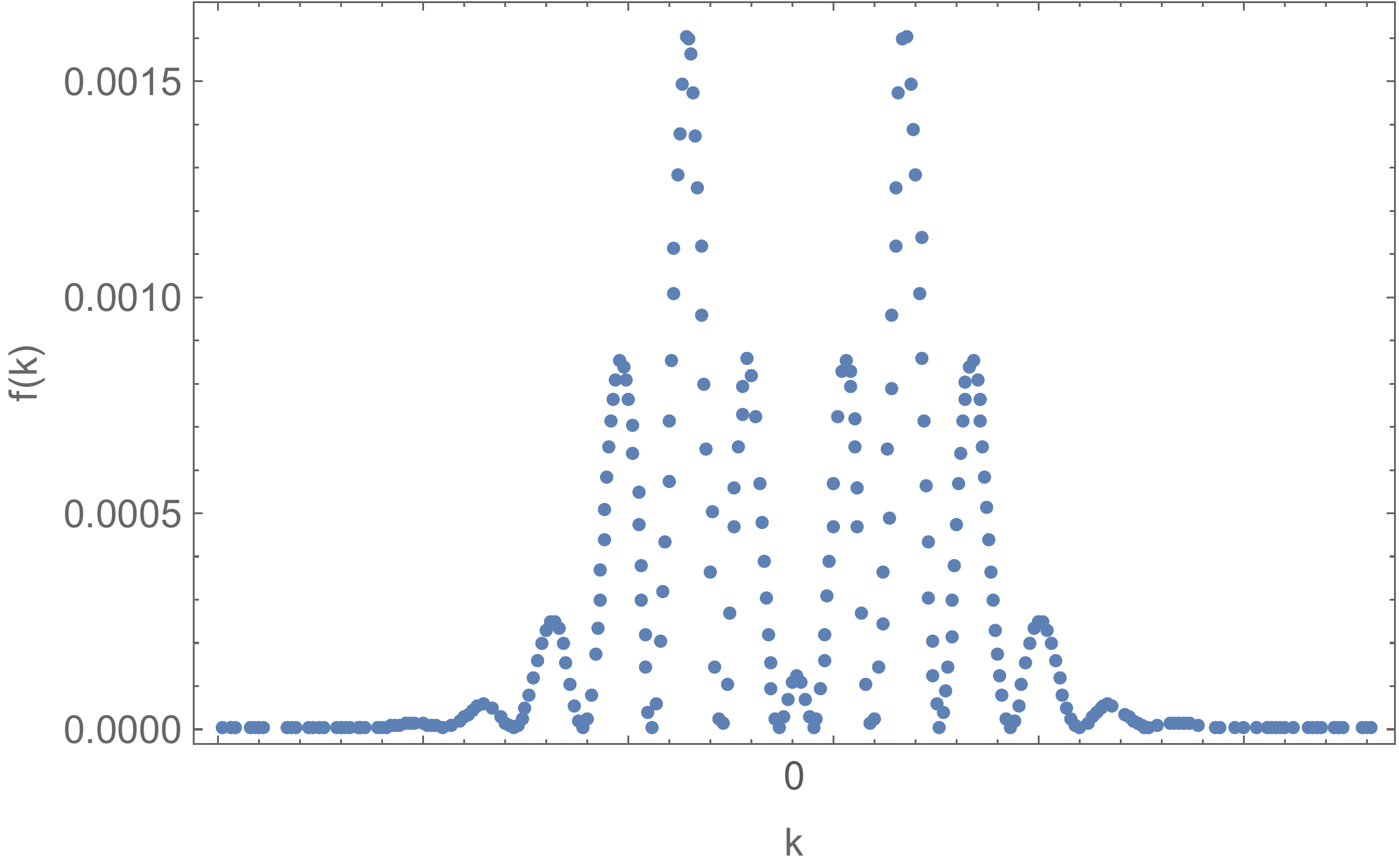}
\end{center}
\caption{Modal fractions for $A=B=1,$ $\protect\mu =1,$ $\protect\eta %
_{1}^{00}=\protect\eta _{2}^{00}=-3.05,$ $\protect\eta _{1}^{11}=\protect%
\eta _{2}^{11}=1.01,$ $\protect\eta _{1}^{01}=\protect\eta _{2}^{01}=1.01,$ $%
\protect\eta _{1}^{10}=\protect\eta _{2}^{10}=1.02,$ $k^{01}=0.03,$ $%
k^{10}=0.02,$ $k^{11}=0.01,$ and $c_{0}=-2.000001.$The plot on the top of
the figure show the case with $k^{00}=-0.058$ and the bottom ones with $%
k^{00}=-0.056$.}
\end{figure}

\begin{figure}[h]
\begin{center}
\includegraphics[width=13cm]{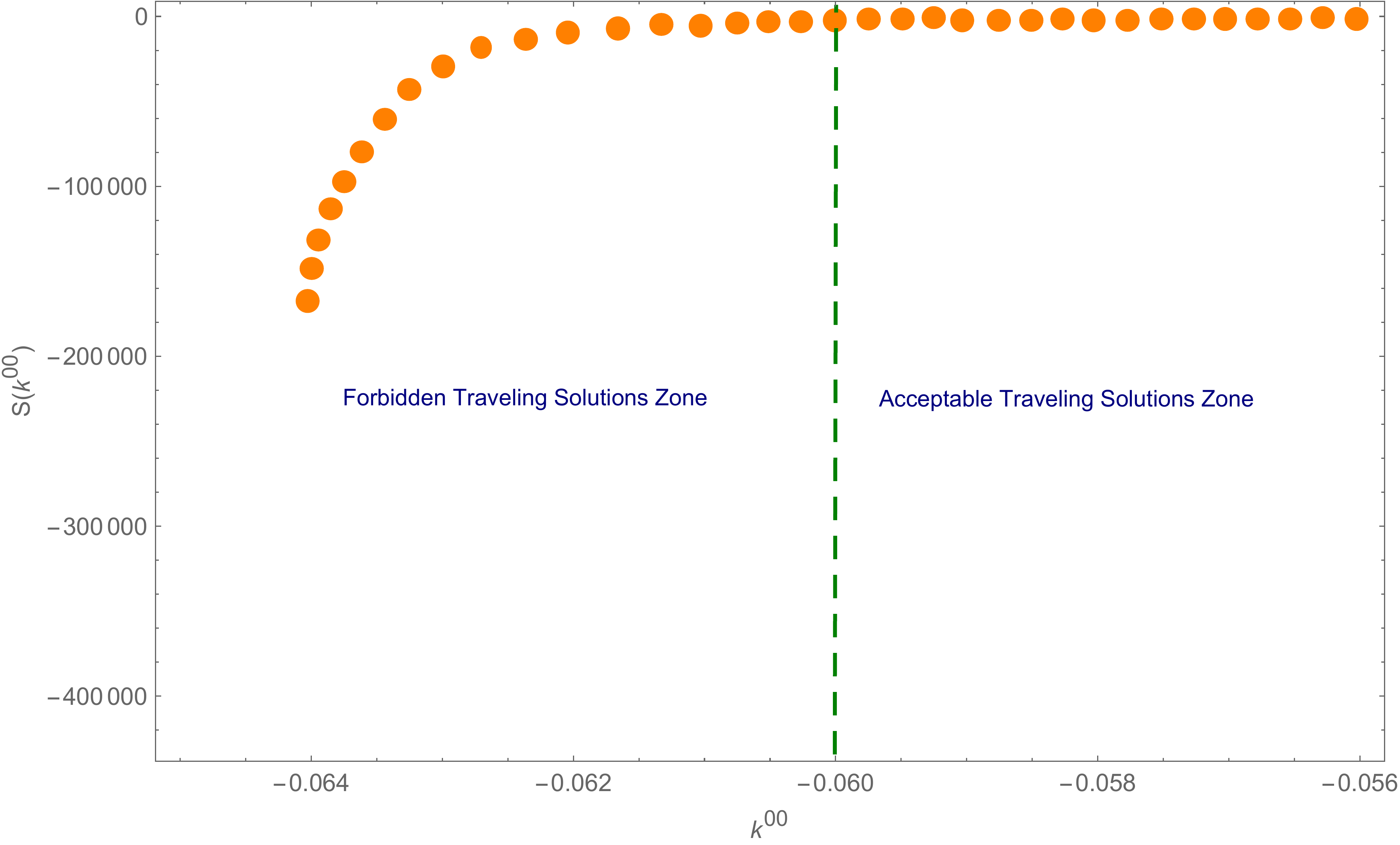}
\end{center}
\caption{ travelling configurational entropy for  $A=B=1,$ $\protect\mu =1,$ $%
\protect\eta _{1}^{00}=\protect\eta _{2}^{00}=-3.05,$ $\protect\eta %
_{1}^{11}=\protect\eta _{2}^{11}=1.01,$ $\protect\eta _{1}^{01}=\protect\eta %
_{2}^{01}=1.01,$ $\protect\eta _{1}^{10}=\protect\eta _{2}^{10}=1.02,$ $%
k^{01}=0.03,$ $k^{10}=0.02,$ $k^{11}=0.01,$ and $c_{0}=-2.000001.$}
\end{figure}

\end{document}